\documentclass[english,aps,reprint,prl]{revtex4-2}
\usepackage[T1]{fontenc}
\usepackage[latin9]{inputenc}
\usepackage{color}
\usepackage{babel}
\usepackage{mathtools}
\usepackage{amsmath}
\usepackage{amssymb}
\usepackage{stmaryrd}
\usepackage[unicode=true,pdfusetitle,
 bookmarks=true,bookmarksnumbered=false,bookmarksopen=false,
 breaklinks=true,pdfborder={0 0 0},pdfborderstyle={},backref=false,colorlinks=true]
 {hyperref}
\hypersetup{
 allcolors=blue}

\makeatletter

\newcommand{\rd}{\mathrm{d}}

\makeatother

\begin{document}
\title{Nonequilibrium Stationary Process and Fluctuation-Dissipation Relations}
\author{Ying-Jen Yang}
\email{yangyj@uw.edu}

\affiliation{Department of Applied Mathematics, University of Washington, Seattle,
WA 98195, USA}
\author{Hong Qian}
\email{hqian@uw.edu}

\affiliation{Department of Applied Mathematics, University of Washington, Seattle,
WA 98195, USA}
\begin{abstract}
A stochastic dynamics has a natural decomposition into a drift capturing mean rate of change and a martingale increment capturing randomness. They are two statistically uncorrelated, but not necessarily independent mechanisms contributing to the overall fluctuations of the dynamics, representing the uncertainties in the past and in the future. A generalized Einstein relation is a consequence solely because the dynamics being stationary; and the Green-Kubo formula reflects a balance between the two mechanisms. Equilibrium with reversibility is characterized by a novel covariance symmetry.
\end{abstract}
\maketitle

\paragraph{Introduction}

Each step, small or large, of a complex motion can be represented by a ``stochastic noise part'' and a ``deterministic drift part'' whose interplay gives rise to the fluctuation-dissipation relation (FDR), a central result in statistical physics. From a dynamical systems point of view, a dissipative drift represents a contraction in the state space (many-to-1 in discrete state space) while a stochastic motion implies divergent moves (1-to-many with probabilities). When these two ``opposing'' tendencies strike a balance and yield a stationary process, the FDR appears.  

There are at least two quantitative manifestations of the above physical picture: On the one hand, the classic FDR in linear response theory, following Onsager's regression hypothesis \cite{onsager31}, is a set of relations between a system's relaxation-after-perturbation near an equilibrium and auto-correlation of spontaneous equilibrium fluctuations \citep{callen_irreversibility_1951,green_markoff_1954,kubo_fluctuation-dissipation_1966,evans_application_2005}.
Extensions of this result to nonequilibrium systems can be found in 
recent work \citep{prost_generalized_2009,seifert_fluctuation-dissipation_2010,altaner_fluctuation-dissipation_2016}. 
On the other hand, the Einstein relation and the Green-Kubo formula (GKF) are between components with different statistical characteristics,
dissipative vs. fluctuating, within a stationary process without a perturbation  \citep{zwanzig,elson}.  
To illustrate, using Langevin's equation for the velocity of a Brownian particle that follows $m(\rd V/\rd t)=-\eta V + \sqrt{2 k_BT \eta}\xi(t)$, $\xi(t)$ being a standard white noise, stationary $\mathbb{E}[\Delta V(t)\Delta V(t+\tau)] =(k_BT/m)e^{-\eta\tau/m}$ where $\Delta V(t)\coloneqq V(t)-\mathbb{E}[ V(t)]$. 
Then,
\begin{equation}
\underbrace{ D= \frac{k_BT} {\eta} }_{\text{ Einstein relation}} \underbrace{ = \int_0^{\infty} \mathbb{E}[\Delta V(t)\Delta V(t+\tau)] \rd\tau }_{\text{Green-Kubo formula }}, 
\label{e1}
\end{equation}
in which $V(t)$ is a stationary process, and   
$D$ is the long-time limit of the mean square displacement of $X(t)$, the integration of $V(t)$:
$\langle (X(t)-X(0))^2 \rangle\sim 2Dt$. In this Letter, we shall refer relations between diffusion and drift as generalized Einstein relation (GER) and relations involving time correlation function as the GKF. Extensions of them to the nonequilibrium realm have also been explored: For a $n$-dimensional linear irreversible Ornstein-Uhlenbeck process, GER takes the form of the Lyapunov matrix equation, connecting to the theory of linear stability and control \citep{keizer,qian_mathematical_2001},
\begin{equation}
  2\mathbf{D}=\boldsymbol{\Xi}\mathbf{B}^{\mathsf{T}}+\mathbf{B}\boldsymbol{\Xi},   
\label{eq: Lyapunov matrix equation}
\end{equation}
where $\boldsymbol{\Xi}$, $\mathbf{B}$, and $\mathbf{D}$ are matrices of the stationary covariance between states, the linear relaxation, and the diffusion. A nonlinear GKF for nonequilibrium steady state in continuous time Markov
processes was also established in Ref. \citep{chen_greenkubo_2006,jiang_mathematical_2004}:
\begin{equation}
 \mathbb{E}\big[D(X_t)\big] = \int_0^{\infty}
  \mathbb{E}\big[b(X_t)b(X_{t+\tau}) \big]\rd\tau,
\label{e3}
\end{equation}
where stationary $X_t$ follows a nonlinear stochastic differential equation with
drift $b(X_t)$ and diffusion $D(X_t)$.

In this Letter, we show that ``noise'' and ``drift'' with distinct statistical characteristics can be used as defining properties for a decomposition of general stationary processes, without additional supposition of Markovian, reversibility, linearity, etc.  
In terms of this decomposition, a GER can be formulated.  The novel formalism clearly illustrates, actually it defines, the GER as the consequence of stationarity of a process in which the stochasticity is balanced by dissipation. 
In the past, the stationary Fokker-Planck equation for a Markov process poses a mathematical relation among three: (i) the stationary distribution, (ii) the dissipative drift, and (iii) the stochastic noise strength \citep{grimmett_probability_2001}.  For most of the applications, one solves (i) based on (ii) and (iii).  Alternatively, given (i) and  (iii), (ii)
admits a general decomposition in terms of the other two \citep{ao_potential_2004,wang_potential_2008,yang_potentials_2021}.  All these previous results are now encompassed in the GER (Eq. \ref{eq: FDR} below) and the GKF in Eq. (\ref{eq: GKR martingale form}), broadly generalizing (\ref{eq: Lyapunov matrix equation}) and (\ref{e3}), respectively.

A key mathematical insight contributing to our result is a general decomposition of stochastic processes discovered by Doob \citep{doob_stochastic_1990,grimmett_probability_2001}:
A stochastic step can always be written as the sum of a ``drift part'' that captures the average increment and a ``noise part'' that captures
the stochasticity. The latter parts from all steps constitute a martingale \citep{grimmett_probability_2001}, which is a process that has no (conditional) gain or loss on average: a fair game. In stochastic thermodynamics, the theory of martingale has played an important role in studying stopping time statistics \citep{neri_statistics_2017,manzano_thermodynamics_2021}. We discover that
the martingale increment is uncorrelated to the past, which leads to a clear-cut of two uncorrelated contributions to the fluctuations of the overall stochastic dynamics, as shown in Eq. (\ref{eq: covariance decomposition}). 


For Markov processes with detailed balance, our GER leads to another, new characterization of equilibrium steady state: By considering a process and its adjoint process, we show that the covariance matrix between the state and drift is symmetric if the process is reversible.  
Various forms of the GKF, as
corollaries, can be derived for the adjoint drift. We can also identify the 1-to-many and many-to-1 features in the dynamics, representing uncertainties in the future and in the past respectively, with the fluctuations of drift and adjoint drift under Doob decompositions of a process and its adjoint.  

All results point to the stationarity being central to FDR.  The search for a similar relationship, between noise and drift, in sweeping dynamics that does not reach stationarity \citep{qian_mathematical_2001} naturally arises.  We shall briefly discuss one class of sweeping processes whose exponentiation becomes a martingale.

\paragraph{Doob Decomposition}

We consider a general discrete time $n$-dimensional ($n$-D) stochastic process, not necessarily Markovian, $\mathbf{X}_{t}\in\mathbb{R}^n$, $t\in\mathbb{N}$.  Continuous time processes can be discussed by considering their infinitesimal $\rd t$ and taking the continuous time limit. We use
$\mathbf{X}_{0:t}$ to denote the entire stochastic trajectory from time $0$ to time $t$.  The change of the value of the state
of the system from time $t$ to $t+1$ has a natural decomposition
by the conditional expectation: 
\begin{align}
\delta\mathbf{X}_{t} & \coloneqq\mathbf{X}_{t+1}-\mathbf{X}_{t}\nonumber \\
 & =\delta\mathbf{A}_{t}\left(\mathbf{X}_{0:t}\right)+\delta\mathbf{M}_{t}\left(\mathbf{X}_{0:t+1}\right)\label{eq: Doob's decomposition}
\end{align}
where \begin{subequations}
\begin{align}
\delta\mathbf{A}_{t} & \coloneqq\mathbb{E}\big[\mathbf{X}_{t+1}|\mathbf{X}_{0:t}\big]-\mathbf{X}_{t}\label{eq: drift def}\\
\delta\mathbf{M}_{t} & \coloneqq\mathbf{X}_{t+1}-\mathbb{E}\big[\mathbf{X}_{t+1}|\mathbf{X}_{0:t}\big].\label{eq: martingale def}
\end{align}
\end{subequations}The first term $\delta\mathbf{A}_{t}$ in the decomposition
is the conditional average change of $\mathbf{X}_{t}$, a function
of entire, non-Markovian $\mathbf{X}_{0:t}$, that captures the average dynamics of $\mathbf{X}_{t}$,
$\mathbb{E}\left[\delta\mathbf{X}_{t}|\mathbf{X}_{0:t}\right]=\delta\mathbf{A}_{t}.$
Hence, the increment $\delta\mathbf{A}_{t}$ is referred as the \emph{drift }of $\mathbf{X}_{t}$\emph{.} The second term $\delta\mathbf{M}_{t}$ captures the 1-to-many randomness in the change of $\mathbf{X}_{t}$, it has a zero (conditional)
mean: 
\begin{equation}
\mathbb{E}\left[\delta\mathbf{M}_{t}|\mathbf{X}_{0:t}\right]=0\text{ and }\mathbb{E}\left[\delta\mathbf{M}_{t}\right]=0.\label{eq: mean zero of delta M}
\end{equation}
The existence of this decomposition of a general process in Eq. (\ref{eq: Doob's decomposition}) into the sum of two processes is known as the \emph{Doob decomposition theorem} \citep{doob_stochastic_1990,grimmett_probability_2001}.

The decomposed process $\mathbf{M}_{t}=\sum_{k=0}^{t-1}\delta\mathbf{M}_{t}$
satisfies
\begin{equation}
\mathbb{E}\left[\mathbf{M}_{t}|\mathbf{X}_{0:s}\right]=\mathbf{M}_{s},\text{ for all }0\le s\le t\label{eq: zero gain in martingale}
\end{equation}
due to the zero conditional gain in every increment. In the theory
of probability, such a process is called a \emph{martingale }\citep{grimmett_probability_2001,shreve_stochastic_2010}\emph{.
}Typical examples of martingales are an unbiased random walk in discrete
time and a Brownian motion in the continuous time.

The zero (conditional) mean properties of the martingale increment
in Eq. (\ref{eq: mean zero of delta M}) implies that $\delta\mathbf{M}_{t}$
is an increment uncorrelated to the past (but not necessarily independent):
for an arbitrary path scalar variable of $\mathbf{X}_{0:t}$, $f\left(\mathbf{X}_{0:t},t\right)$,
the covariance between $f$ and any component of $\delta\mathbf{M}_{t}$, say the $i$th one
denoted as $\delta M_{t}^{(i)}$, is zero:
\begin{equation}
{\rm Co}\mathbb{V}\left[f\left(\mathbf{X}_{0:t},t\right),\delta M_{t}^{(i)}  \right]=0.
\label{eq: uncorrelated increment of M}
\end{equation}
This leads to the following two important results. To present them
in a more concise way, we will use $\left\llbracket \mathbf{u},\mathbf{w}\right\rrbracket $
to denote the covariance matrix between two vector random variables
$\mathbf{u}$ and $\mathbf{w}$ in this paper. Specifically,
the $i,j$ component of $\left\llbracket \mathbf{u},\mathbf{w}\right\rrbracket $
is ${\rm Co}\mathbb{V}\left[u_{i},w_{j}\right].$ 

First, the martingale increments at different times are
uncorrelated $\left\llbracket \delta\mathbf{M}_{t},\delta\mathbf{M}_{s}\right\rrbracket =\mathbf{0}.$
This leads to the fact that a martingale has an ever-increasing, additive fluctuation,
\begin{equation}
\left\llbracket \mathbf{M}_{t},\mathbf{M}_{t}\right\rrbracket =\sum_{k=0}^{t-1}\left\llbracket \delta\mathbf{M}_{k},\delta\mathbf{M}_{k}\right\rrbracket .\label{eq: martingale variance increment}
\end{equation}
The scalar version of this, $\mathbb{V}\left[M_{t}\right]=\sum_{k=0}^{t-1}\mathbb{V}\left[\delta M_{k}\right]$,
is a discrete-time analogue of It\^{o} isometry \citep{shreve_stochastic_2010}
and is, in a sense, more general than It\^{o} isometry since the
martingale in It\^{o} isometry is the Brownian motion which has independent
increments whereas Eq. (\ref{eq: martingale variance increment})
doesn't require independency in the increments.

Second, the uncertainty of increment $\delta\mathbf{X}_{t}$
actually has two uncorrelated sources identified by the Doob decomposition
in Eq. (\ref{eq: Doob's decomposition}),
\begin{equation}
\left\llbracket \delta\mathbf{X}_{t},\delta\mathbf{X}_{t}\right\rrbracket =\left\llbracket \delta\mathbf{A}_{t},\delta\mathbf{A}_{t}\right\rrbracket +\left\llbracket \delta\mathbf{M}_{t},\delta\mathbf{M}_{t}\right\rrbracket.\label{eq: covariance decomposition}
\end{equation}
The two sources of the uncertainty
in $\delta\mathbf{X}_{t}$ are rather disjoint conceptually. Since
$\delta\mathbf{A}_{t}$ is a function of the past path $\mathbf{X}_{0:t}$,
the fluctuation of $\delta\mathbf{A}_{t}$ is really about the uncertainty
of the past. On the contrary, the uncertainty of $\delta\mathbf{M}_{t}$
is about the fluctuation in the conditional mapping from $\mathbf{X}_{t}$
to $\mathbf{X}_{t+1}$. If the conditional mapping is deterministic,
then $\delta\mathbf{M}_{t}=0$ but $\left\llbracket \delta\mathbf{A}_{t},\delta\mathbf{A}_{t}\right\rrbracket $
could still be nonzero if there is uncertainty in the initial condition
and/or the past state. 

The two results shown above are valid for general processes. Assumptions
such as Markovian, stationarity, or detailed balance are not needed.
In fact, the Doob decomposition can also be applied to an arbitrary
path variable $\mathcal{U}_{t}\left(\mathbf{X}_{0:t}\right)$. The
decomposition then becomes $\delta\mathcal{U}_{t}=\delta A_{\mathcal{U}_{t}}+\delta M_{\mathcal{U}_{t}}$
where $\delta A_{\mathcal{U}_{t}}\coloneqq\mathbb{E}\left[\mathcal{U}_{t+1}|\mathbf{X}_{0:t}\right]-\mathcal{U}_{t}$
and $\delta M_{\mathcal{U}_{t}}\coloneqq\mathcal{U}_{t+1}-\mathbb{E}\left[\mathcal{U}_{t+1}|\mathbf{X}_{0:t}\right].$
The results presented above still hold.

We note here that two special classes of processes can be identified
with the Doob decomposition and are the natural extension of a martingale.
If the drift of a scalar process is always non-negative, then the
process is called a \emph{submartingale. }If the drift of a scalar
process is always non-positive, then the process is called a \emph{supermartingale.}
An important example of \emph{submartingale }in stochastic thermodynamics\emph{
}is the housekeeping heat $\mathcal{Q}_{\mathrm{hk}}$. In fact, the
housekeeping heat $\mathcal{Q}_{\mathrm{hk}}$ belongs to a special
class of submartingale where $\exp\left(-\mathcal{Q}_{{\rm hk}}\right)$
becomes a martingale \citep{pigolotti_generic_2017,chetrite_martingale_2019,yang_unified_2020}.
Other types of entropy production in stochastic thermodynamics has
a non-negative average drift but their drifts are not guaranteed to
be non-negative before expectation, and are in general not submartingale
\citep{yang_unified_2020}.

\paragraph{Generalized Einstein relation}

If the process $\mathbf{X}_{t}$ reaches a steady state, the probability
distribution of state no longer changes in time. For
those stationary $\mathbf{X}_{t}$, all its cumulants will be fixed
in time. The average state of $\mathbf{X}_{t}$ would be constant
in time, meaning that the average drift of the observable would be
zero at the steady state $\mathbb{E}_{*}\left[\delta\mathbf{A}_{t}\right]=0$
where $\mathbb{E}_{*}\left[\cdot\right]$ means expectation for the
stationary process. For the evolution of the covariance matrix, we have
\begin{align}
\delta\left\llbracket \mathbf{X}_{t},\mathbf{X}_{t}\right\rrbracket  & \coloneqq\left\llbracket \mathbf{X}_{t+1},\mathbf{X}_{t+1}\right\rrbracket -\left\llbracket \mathbf{X}_{t},\mathbf{X}_{t}\right\rrbracket \nonumber \\
 & =\left\llbracket \delta\mathbf{A}_{t},\mathbf{X}_{t}\right\rrbracket +\left\llbracket \delta\mathbf{A}_{t},\mathbf{X}_{t}\right\rrbracket ^{\mathsf{T}}+\left\llbracket \delta\mathbf{X}_{t},\delta\mathbf{X}_{t}\right\rrbracket \label{eq: covariance evolution}
\end{align}
where $\mathsf{T}$ denotes the transpose of a matrix.
This shows that the covariance of states always have a source given by the fluctuation of transitions. 
Stationarity of $\mathbf{X}_{t}$
 is achieved by the balance between the drift and the fluctuation
such that  $\delta\left\llbracket \mathbf{X}_{t},\mathbf{X}_{t}\right\rrbracket =\mathbf{0}$, 
\begin{equation}
\left\llbracket \delta\mathbf{A}_{t},\mathbf{X}_{t}\right\rrbracket _{*}+\left\llbracket \delta\mathbf{A}_{t},\mathbf{X}_{t}\right\rrbracket _{*}^{\mathsf{T}}=-\left\llbracket \delta\mathbf{X}_{t},\delta\mathbf{X}_{t}\right\rrbracket _{*}\label{eq: FDR}
\end{equation}
where $\left\llbracket \mathbf{u},\mathbf{w}\right\rrbracket _{{\rm *}}$
denotes covariance matrix of $\mathbf{u}$ and $\mathbf{w}$ when
the process is stationary.  This shows that the symmetric part of
$\left\llbracket \delta\mathbf{A}_{t},\mathbf{X}_{t}\right\rrbracket _{*}$
is negative-definite. In physics, the drift $\delta\mathbf{A}_{t}$
in a stable process is considered as the dissipation of the dynamics.
Eq. (\ref{eq: FDR}) is thus a generally-valid \emph{Einstein relation
}(GER) implied by stationarity. In a scalar process, Eq. (\ref{eq: FDR})
reduces to $2{\rm Co}\mathbb{V}_{*}\left(\delta A_{t},X_{t}\right)=-\mathbb{V}_{*}\left[\delta X_{t}\right]$
which shows that the drift $\delta A_{t}$, on average, has an opposite
sign of the value of $X_{t}$ as a ``dissipation'' to $X_{t}$.
 In fact, the result can be extended to arbitrary observable of the
process, $U_{t}\left(\mathbf{X}_{t}\right)$, \emph{e.g. }the energy
of the system. The Doob decomposition of its dynamics becomes $\delta U_{t}=\delta A_{U_{t}}+\delta M_{U_{t}}$
where $\delta A_{U_{t}}\coloneqq\mathbb{E}\left[U_{t+1}|\mathbf{X}_{0:t}\right]-U_{t}$
and $\delta M_{U_{t}}\coloneqq U_{t+1}-\mathbb{E}\left[U_{t+1}|\mathbf{X}_{0:t}\right]$.
The GER then becomes $2{\rm Co}\mathbb{V}_{*}\left(\delta A_{U_{t}},U_{t}\right)=-\mathbb{V}_{*}\left[\delta U_{t}\right]$
which clearly portraits a relation between fluctuation $\mathbb{V}_{*}\left[\delta U_{t}\right]$
and the energy dissipation $\delta A_{U_{t}}$.

In Ornstein-Uhlenbeck (OU) processes described by the stochastic differential
equation ${\rm d}\mathbf{X}_{t}=-\mathbf{B}\mathbf{X}_{t}{\rm d}t+\boldsymbol{\Gamma}{\rm d}\mathbf{W}_{t}$
where $\mathbf{W}_{t}$ is the $n$-D Brownian motion, we have $\delta\mathbf{A}_{t}=-\mathbf{B}\mathbf{X}_{t}{\rm d}t$
and $\left\llbracket {\rm d}\mathbf{X}_{t},{\rm d}\mathbf{X}_{t}\right\rrbracket _{*}=2\mathbf{D}{\rm d}t$ where $\mathbf{D}=\boldsymbol{\Gamma}\boldsymbol{\Gamma}^{\mathsf{T}}/2$
is the diffusion matrix.
Eq. (\ref{eq: FDR}) then reduces to the Einstein relation for linear
systems given by the Lyapunov matrix equation in Eq. (\ref{eq: Lyapunov matrix equation}) \citep{qian_mathematical_2001}. 

The GER in Eq. (\ref{eq: FDR}) is valid for any stationary processes.
Conditions such as Markovian or detailed balance are not needed. It
is a necessary condition for the stationarity of fluctuation and covariance.
We note that the stationarity of $\mathbf{X}_{t}$ actually requires
all of its cumulants to be fixed in time. Thus, any martingale, supermartingale,
or submartingale will not satisfy the GER since a martingale has an
ever-increasing variance and both supermartingale and submartingale
have a monotonic drift. 

\paragraph{Green-Kubo Formula}

Eqs. (\ref{eq: covariance evolution}) and (\ref{eq: FDR}) further
allow us to derive a general Green-Kubo formula (GKF) for a stationary
processes \citep{kubo_fluctuation-dissipation_1966,jiang_mathematical_2004,chen_greenkubo_2006}.
We note that for $t\ge0$, we have
\begin{equation}
\left\llbracket \delta\mathbf{A}_{t},\delta\mathbf{A}_{0}\right\rrbracket =\left\llbracket \mathbf{X}_{t+1},\delta\mathbf{A}_{0}\right\rrbracket -\left\llbracket \mathbf{X}_{t},\delta\mathbf{A}_{0}\right\rrbracket .\label{eq: dAt dAo}
\end{equation}
Therefore, 
\[
\sum_{k=0}^{\infty}\left\llbracket \delta\mathbf{A}_{k},\delta\mathbf{A}_{0}\right\rrbracket =\left\llbracket \mathbf{X}_{\infty},\delta\mathbf{A}_{0}\right\rrbracket -\left\llbracket \mathbf{X}_{0},\delta\mathbf{A}_{0}\right\rrbracket .
\]
Assuming that $\mathbf{X}_{t}$ has a finite correlation time, the
first term is zero. Then, by applying Eq. (\ref{eq: FDR}) to the
above equation, we get a general GKF that relates auto-correlation
of the dissipative drift $\delta\mathbf{A}_{t}$ to the fluctuation
of $\delta\mathbf{M}_{t}$ at steady state, 
\begin{equation}
\left\llbracket \delta\mathbf{X}_{t},\delta\mathbf{X}_{t}\right\rrbracket _{*}=\sum_{k=0}^{\infty}\left\{\left\llbracket \delta\mathbf{A}_{k},\delta\mathbf{A}_{0}\right\rrbracket _{*}+\left\llbracket \delta\mathbf{A}_{k},\delta\mathbf{A}_{0}\right\rrbracket _{*}^{\mathsf{T}}\right\}
\label{eq: GKR}
\end{equation}
which becomes
\begin{equation}
\left\llbracket \delta\mathbf{M}_{t},\delta\mathbf{M}_{t}\right\rrbracket _{*}=\sum_{k=-\infty}^{\infty}\left\llbracket \delta\mathbf{A}_{k},\delta\mathbf{A}_{0}\right\rrbracket _{*}\label{eq: GKR martingale form}
\end{equation}
by using Eq. (\ref{eq: covariance decomposition}) and stationarity.
Eq. (\ref{eq: GKR martingale form}) shows that the GKF is really
a relation between the drift $\delta\mathbf{A}_{t}$ and the martingale
increment $\delta\mathbf{M}_{t}$. The results for continuous time
processes derived in the past \citep{chen_greenkubo_2006,jiang_mathematical_2004}
lost this important insight since in the continuous time processes
considered have $\delta\mathbf{A}_{t}=O({\rm d}t)$ and $\delta\mathbf{M}_{t}=O(\sqrt{{\rm d}t})$,
which makes the covariance of the drift higher order. \textcolor{red}{}

\paragraph{Adjoint Processes and Adjoint Drift}

The Doob decomposition shown in Eq. (\ref{eq: Doob's decomposition})
is with respect to the forward probability measure $\mathbb{P}.$
In Markov processes with steady states, we can consider the Doob decomposition
given by the adjoint probability measure $\mathbb{P}^{\dagger}$
(\emph{i.e.,} the decomposition in the adjoint process) where the
transition probability is given by 
\[
\mathbb{P}^{\dagger}\left\{ \mathbf{X}_{t+1}=\mathbf{y}|\mathbf{X}_{t}=\mathbf{x}\right\} =\frac{\pi\left(\mathbf{y}\right)}{\pi\left(\mathbf{x}\right)}\mathbb{P}\left\{ \mathbf{X}_{t+1}=\mathbf{x}|\mathbf{X}_{t}=\mathbf{y}\right\} .
\]
The adjont Doob decomposition is then 
\begin{equation}
\delta\mathbf{X}_{t}=\delta\mathbf{A}_{t}^{\dagger}+\delta\mathbf{M}_{t}^{\dagger}\label{eq: adjoint Doob decomposition}
\end{equation}
with $\delta\mathbf{A}_{t}^{\dagger}\coloneqq\mathbb{E}^{\dagger}\left[\mathbf{X}_{t+1}|\mathbf{X}_{t}\right]-\mathbf{X}_{t}$
and $\delta\mathbf{M}_{t}^{\dagger}\coloneqq\mathbf{X}_{t+1}-\mathbb{E}^{\dagger}\left[\mathbf{X}_{t+1}|\mathbf{X}_{t}\right].$
The covariance between $\mathbf{X}_{t}$ and $\delta\mathbf{A}_{t}$
and the covariance between $\mathbf{X}_{t}$ and $\delta\mathbf{A}_{t}^{\dagger}$
are only subject to a transpose at steady state, 
\begin{equation}
\left\llbracket \mathbf{X}_{t},\delta\mathbf{A}_{t}\right\rrbracket _{*}=\left\llbracket \mathbf{X}_{t},\delta\mathbf{A}_{t}^{\dagger}\right\rrbracket _{*}^{\mathsf{T}}.\label{eq: covariance X A and X A dagger}
\end{equation}
This gives us a neater expression of the GER in Eq. (\ref{eq: FDR}),
\begin{equation}
\left\llbracket \mathbf{X}_{t},\delta\mathbf{A}_{t}+\delta\mathbf{A}_{t}^{\dagger}\right\rrbracket _{*}=-\left\llbracket \delta\mathbf{X}_{t},\delta\mathbf{X}_{t}\right\rrbracket _{*}.\label{eq: GFDR with adjoint}
\end{equation}
This allows one to show that for reversible (detailed balanced) systems
where the forward and the adjoint processes are the same, we have
$\delta\mathbf{A}_{t}=\delta\mathbf{A}_{t}^{\dagger}$ and thus 
\begin{equation}
\left\llbracket \mathbf{X}_{t},\delta\mathbf{A}_{t}\right\rrbracket _{*}=-\frac{1}{2}\left\llbracket \delta\mathbf{X}_{t},\delta\mathbf{X}_{t}\right\rrbracket _{*}\label{eq: GFDR for DB systems}
\end{equation}
 which shows that the covariance matrix $\left\llbracket \mathbf{X}_{t},\delta\mathbf{A}_{t}\right\rrbracket _{*}$
is symmetric and negative-definite. This gives yet another characterization
of detailed balance and is the generalization of $\mathbf{B}\boldsymbol{\Xi}$
being symmetric for reversible OU processes \citep{qian_mathematical_2001}.
We note that the GKF also have two sibling expressions in terms of
the adjoint drift for continuous time Markov chain \citep{chen_greenkubo_2006}.
One can derive \begin{subequations}\label{eq: siblings of GKR}
\begin{align}
\left\llbracket \delta\mathbf{X}_{t},\delta\mathbf{X}_{t}\right\rrbracket _{*} & =\sum_{k=0}^{\infty}\left\llbracket \delta\mathbf{A}_{k},\delta\mathbf{A}_{0}^{\dagger}\right\rrbracket _{*}+\left\llbracket \delta\mathbf{A}_{k},\delta\mathbf{A}_{0}^{\dagger}\right\rrbracket _{*}^{\mathsf{T}} \label{eq: A Adagger}\\
 & =\sum_{k=0}^{\infty}\left\llbracket \delta\mathbf{A}_{k}^{\dagger},\delta\mathbf{A}_{0}^{\dagger}\right\rrbracket _{*}+\left\llbracket \delta\mathbf{A}_{k}^{\dagger},\delta\mathbf{A}_{0}^{\dagger}\right\rrbracket _{*}^{\mathsf{T}}\label{eq: A dagger A dagger}
\end{align}
\end{subequations}by a similar approach \footnote{The derivation is by formulating the drift of a arbitrary function
of the process $f\left(\mathbf{X}_{t}\right)$ as an operator $\mathcal{A}$
operating on $f$: $\mathcal{A}f\left(\mathbf{X}_{t}\right)\coloneqq\mathbb{E}\left[f(\mathbf{X}_{t+1})|\mathbf{X}_{0:t}\right]-f(\mathbf{X}_{t}).$
The adjoint drift is $\mathcal{A}^{\dagger}f\left(\mathbf{X}_{t}\right)\coloneqq\mathbb{E}^{\dagger}\left[f(\mathbf{X}_{t+1})|\mathbf{X}_{t}\right]-f(\mathbf{X}_{t})$.
A generalization to Eq. (\ref{eq: dAt dAo}) is $\left\llbracket \mathcal{A}f\left(\mathbf{X}_{t}\right),g\left(\mathbf{X}_{0}\right)\right\rrbracket =\left\llbracket f\left(\mathbf{X}_{t+1}\right),g\left(\mathbf{X}_{0}\right)\right\rrbracket -\left\llbracket f\left(\mathbf{X}_{t}\right),g\left(\mathbf{X}_{0}\right)\right\rrbracket .$
One can then prove Eq. (\ref{eq: GKR}) by using $f=X_{t}^{(i)}$
and $g=\delta A_{0}^{(j)},$ prove Eq.
(\ref{eq: A Adagger}) by using $f=X_{t}^{(i)}$
and $g=\delta A_{0}^{\dagger,(j)}$ for
Markov processes, and prove Eq. (\ref{eq: A dagger A dagger}) by
using $f=\delta A_{0}^{\dagger,(j)}$ and
$g=X_{0}^{(j)}$ for Markov processes and
$\mathbb{E}_{*}\left[f\left(\mathbf{X}_{t}\right)\mathcal{A}g\left(\mathbf{X}_{t}\right)\right]=\mathbb{E}_{*}\left[g\left(\mathbf{X}_{t}\right)\mathcal{A}^{\dagger}f\left(\mathbf{X}_{t}\right)\right]$.
All three relations rely on the asymptotic independence between $\mathbf{X}_{\infty}$
and $\mathbf{X}_{0}$.}. 

\paragraph{GER in Continuous Markov processes}

In a continuous process described by stochastic differential equations,
\begin{equation}
{\rm d}\mathbf{X}_{t}=\left(\mathbf{b}+\nabla\cdot\mathbf{D}\right){\rm d}t+\boldsymbol{\Gamma}{\rm d}\mathbf{W}_{t},\label{eq: SDE}
\end{equation}
where $\mathbf{D}=\boldsymbol{\Gamma}\boldsymbol{\Gamma}^{\mathsf{T}}/2$
is the diffusion matrix and $\mathbf{W}_{t}$ is the Brownian motion,
we have $\delta\mathbf{A}_{t}=\left(\mathbf{b}+\nabla\cdot\mathbf{D}\right){\rm d}t$
and $\delta\mathbf{M}_{t}=\boldsymbol{\Gamma}{\rm d}\mathbf{W}_{t}$.
The vector field decomposition $\mathbf{b}=-\mathbf{D}\nabla\Phi-\mathbf{Q}\nabla\Phi+\nabla\times\mathbf{Q}$
discussed in Ref. \citep{yang_potentials_2021} provides an alternative
proof for the GER in continuous processes. By using integration by
part, $\mathbf{Q}=-\mathbf{Q}^{\mathsf{T}}$ and $\mathbf{D}=\mathbf{D}^{\mathsf{T}}$,
the covariance $\left\llbracket \mathbf{X}_{t},\delta\mathbf{A}_{t}\right\rrbracket _{*}=\left\llbracket \mathbf{X}_{t},\mathbf{b}+\nabla\cdot\mathbf{D}\right\rrbracket _{*}{\rm d}t$
can be further rewritten as 
\begin{align}
\left\llbracket \mathbf{X}_{t},\delta\mathbf{A}_{t}\right\rrbracket _{\mathrm{*}} & =\mathbb{E}_{*}\left[\mathbf{Q}-\mathbf{D}\right]{\rm d}t.\label{eq: Q-D FDR}
\end{align}
This shows that $\left\llbracket \mathbf{X}_{t},\delta\mathbf{A}_{t}\right\rrbracket _{*}+\left\llbracket \mathbf{X}_{t},\delta\mathbf{A}_{t}\right\rrbracket _{*}^{\mathsf{T}}=\mathbb{E}_{*}\left[\mathbf{Q}-\mathbf{D}\right]{\rm d}t+\mathbb{E}_{*}\left[-\mathbf{Q}-\mathbf{D}\right]{\rm d}t=-2\mathbb{E}_{*}\left[\mathbf{D}\right]{\rm d}t$,
which is exactly the continuous-time version of Eq. (\ref{eq: GKR}).
With $\mathbf{Q}$ understood as the cycle velocity in continuous
processes \citep{yang_bivectorial_2021,yang_potentials_2021}, Eq.
(\ref{eq: Q-D FDR}) actually presents an interesting relation between
average cycle velocity $\mathbb{E}_{*}\left[\mathbf{Q}\right]$, average
diffusion matrix $\mathbb{E}_{*}\left[\mathbf{D}\right]$, and covariance
between system state and its conditional drift velocity $\left\llbracket \mathbf{X}_{t},\frac{\delta\mathbf{A}_{t}}{{\rm d}t}\right\rrbracket _{*}$.
We note that $\mathbf{Q}=\mathbf{0}$ corresponds to detailed balanced
systems. This again implies the covariance $\left\llbracket \mathbf{X}_{t},\delta\mathbf{A}_{t}\right\rrbracket _{*}$
is a symmetric and negative-definite matrix in detailed balanced systems.

\paragraph{Reversed Decomposition}

The Doob decomposition decomposes the dynamics ${\rm \delta}\mathbf{X}_{t}$
into a drift part $\delta\mathbf{A}_{t}$ and a martingale part $\delta\mathbf{M}_{t}$. 
One of our key results is that the uncertainty of the dynamics has a resulting uncorrelated
decomposition into the fluctuation of the past and the fluctuation
of the 1-to-many mapping marching toward the future as shown in Eq.
(\ref{eq: covariance decomposition}). Here, we show another decomposition
that relates the fluctuation of the dynamics to the many-to-1 uncertainty
in the dynamics.

We can decompose $\delta\mathbf{X}_{t}$ by conditioning on the state
one step in the future,
\begin{align}
\delta\mathbf{X}_{t} & =\delta\mathbf{R}_{t}+\delta\mathbf{N}_{t}\label{eq: reversed decomposition}
\end{align}
where \begin{subequations}
\begin{align}
\delta\mathbf{R}_{t} & =\mathbf{X}_{t+1}-\mathbb{E}\left[\mathbf{X}_{t}|\mathbf{X}_{t+1}\right]\label{eq: reversed drift}\\
\delta\mathbf{N}_{t} & =\mathbb{E}\left[\mathbf{X}_{t}|\mathbf{X}_{t+1}\right]-\mathbf{X}_{t}.\label{eq: delta N}
\end{align}
\end{subequations} This is also an uncorrelated decomposition, $\left\llbracket \delta\mathbf{R}_{t},\delta\mathbf{N}_{t}\right\rrbracket =\mathbf{0}$,
which leads us to another fluctuation decomposition,
\begin{equation}
\left\llbracket \delta\mathbf{X}_{t},\delta\mathbf{X}_{t}\right\rrbracket =\left\llbracket \delta\mathbf{R}_{t},\delta\mathbf{R}_{t}\right\rrbracket +\left\llbracket \delta\mathbf{N}_{t},\delta\mathbf{N}_{t}\right\rrbracket .\label{eq: backward fluctuation decomposition}
\end{equation}
$\delta\mathbf{R}_{t}$ can be thought of as the backward
drift and $\delta\mathbf{N}_{t}$ is a quantification of the many-to-1
uncertainty in the dynamics.
Therefore, Eq. (\ref{eq: backward fluctuation decomposition}) and Eq. (\ref{eq: covariance decomposition}) together link the many-to-1 uncertainty and 1-to-many uncertainty with forward and backward drift,
\begin{subequations}
\begin{align}
\left\llbracket \delta\mathbf{X}_{t},\delta\mathbf{X}_{t}\right\rrbracket = & \left\llbracket \delta\mathbf{A}_{t},\delta\mathbf{A}_{t}\right\rrbracket +\left\llbracket \delta\mathbf{M}_{t},\delta\mathbf{M}_{t}\right\rrbracket \label{eq: AA MM}\\
= & \left\llbracket \delta\mathbf{R}_{t},\delta\mathbf{R}_{t}\right\rrbracket +\left\llbracket \delta\mathbf{N}_{t},\delta\mathbf{N}_{t}\right\rrbracket .\label{eq: Ad Ad NN}
\end{align}
\end{subequations}

For Markov processes, the backward drift becomes the adjoint drift at steady state, $\delta\mathbf{R}_{t}=-\delta\mathbf{A}_{t}^{\dagger}$.
In fact, the decomposition in Eq. (\ref{eq: reversed decomposition})
is actually from the Doob decomposition of the reversed process $\mathbf{Z}_{t}=\mathbf{X}_{-t}$.
The past and the future is conditionally independent
if conditioned on the present in Markov processes,
\begin{equation}
P_{\mathbf{X}_{0:t}|\mathbf{X}_{t+1:\infty}}\left(\mathbf{x}_{0:t}|\mathbf{x}_{t+1:\infty}\right)=P_{\mathbf{X}_{0:t}|\mathbf{X}_{t+1}}\left(\mathbf{x}_{0:t}|\mathbf{x}_{t+1}\right).\label{eq: conditional independency in MP}
\end{equation}
This means that the conditional expectation $\mathbb{E}\left[\mathbf{X}_{t}|\mathbf{X}_{t+1}\right]$
in Eq. (\ref{eq: reversed decomposition}) is the same as conditioning
the whole future, $\mathbb{E}\left[\mathbf{X}_{t}|\mathbf{X}_{t+1:\infty}\right].$
The decomposed process is thus
\begin{equation}
\mathbf{N}_{t}=\sum_{k=0}^{t-1}\delta\mathbf{N}_{k}=\sum_{k=0}^{t-1}\left(\mathbb{E}\left[\mathbf{X}_{t}|\mathbf{X}_{t+1:\infty}\right]-\mathbf{X}_{t}\right),\label{eq: backward martingale}
\end{equation}
a \emph{reversed martingale}, or called \emph{backward martingale} in mathematics \citep{grimmett_probability_2001}, for Markov
processes.

\paragraph{Exponential Martingale and Work-like Observable}

The GER discussed above is for general stationary processes. It, however,
excludes any martingale, submartingale, and supermartingale. Here, we present
that a specific type of submartingale in continuous processes actually
have a fluctuation-drift relation. We call a process  $\mathcal{E}_{t}$ an exponential
martingale if $e^{-\mathcal{E}_{t}}$ becomes a martingale.
Stochastic calculus tells us that the drift $\mu_{t}$
and the fluctuation level $\sigma_{t}$ in the stochastic differential
equation of $\mathcal{E}_{t}$, ${\rm d}\mathcal{E}_{t}=\mu_{t}{\rm d}t+\sigma_{t}{\rm d}W_{t}$,
is related by 
\begin{equation}
2\mu_{t}=\sigma_{t}^{2}.\label{eq: fluctuation drift relation in exponential martingale}
\end{equation}
This shows that the process $\mathcal{E}_{t}$ is a submartingale
and has a fluctuation-drift relation induced by the requirement of
being a martingale after exponentiation. A famous example for $\mathcal{E}_{t}$
in stochastic thermodynamics is the housekeeping heat $\mathcal{Q}_{{\rm hk}}$
\citep{pigolotti_generic_2017,chetrite_martingale_2019}. 

If an exponential martingale is also work-like, \emph{i.e.} $\rd \mathcal{E}_t$ can be expressed as $\mathbf{f}\circ{\rm d}\mathbf{X}_{t}$ where
$\circ$ denotes Stratonovich midpoint integration, it satisfies a Cauchy-Schwarz inequality given the inner product
$\left\langle \mathbf{u},\mathbf{w}\right\rangle \coloneqq \mathbb{E}\left[\mathbf{u}\cdot\mathbf{D}^{-1}\mathbf{w}\right]$
\citep{dechant_entropic_2018,ito_stochastic_2020},
\begin{equation}
\left(\mathbb{E}\left[\frac{{\rm d}\mathcal{E}_{t}}{{\rm d}t}\right]\right)^{2}\le\mathbb{E}\left[\frac{{\rm d}\mathcal{S}_{{\rm tot}}}{{\rm d}t}\right]\frac{\mathbb{V}\left[{\rm d}\mathcal{E}_{t}\right]}{2{\rm d}t,}\label{eq: Cauchy-Schwarz}
\end{equation}
which is an instantaneous version of the recently-celebrated thermodynamic uncertainty relation \citep{barato_thermodynamic_2015,horowitz_thermodynamic_2020}.
Then, by Eq. (\ref{eq: fluctuation drift relation in exponential martingale}) and time integration, one gets
\begin{equation}
\mathbb{E}\left[\mathcal{E}_{t}\right]\le\mathbb{E}\left[\mathcal{S}_{{\rm tot}}\right].\label{eq: average work if exponential martingale}
\end{equation}
The total entropy production is the upper bound of any work-like exponential martingale.
The force $\mathbf{f}$ should satisfy $\mathbf{f}\cdot\mathbf{D}\mathbf{f}=\mathbf{b}\cdot\mathbf{f}+\nabla\cdot\left(\mathbf{D}\mathbf{f}\right)$ for a work-like process to be an exponential martingale.
Examples for
such a process $\mathcal{E}_{t}$ include the housekeeping heat $\mathcal{Q}_{{\rm hk}}$
and the heat dissipation $\mathcal{Q}$ if the vector field $\mathbf{b}$
is divergence-free.

\paragraph{Conclusion}
In this Letter, we summarize and extend the Einstein relation and the Green-Kubo formula to non-equilibrium, non-linear and non-Markovian systems in a covariance formalism. Two underlying mechanisms contributing to a stochastic change of a system's state were identified: a ``deterministic'' drift summarizing the past and a noise representing the stochasticity of one-step toward the future. 
Stationarity of the process requires a dissipative drift to balance out the fluctuation generated by noise, which is the origin of our general fluctuation-dissipation relations.
Reversibility and Markovian are not needed but can impose a further symmetry in the covariance between the state and the drift.
General relations between the fluctuation and the drift of sweeping dynamics remains to be investigated.
Generally speaking, a symmetry is needed to dictate a fluctuation-drift relation.

\begin{acknowledgments}
H. Q. thanks the support from the Olga Jung Wan Endowed Professorship.
\end{acknowledgments}

\bibliography{FDR}

\end{document}